\documentclass[12pt]{article}
\usepackage{epsfig}
\newcommand{\be}{\begin{equation}}
\newcommand{\ee}{\end{equation}}
\newcommand{\bea}{\begin{eqnarray}}
\newcommand{\eea}{\end{eqnarray}}

\begin{document}

\begin{flushright}
\end{flushright}

\begin{center}

{\Large \bf Controlled Antihydrogen Propulsion for \\
NASA's Future in Very Deep Space}\footnote
{Email addresses: mmn@lanl.gov, michael.holzscheiter@cern.ch,
turyshev@jpl.nasa.gov}

\vspace{2ex}

{\bf Michael Martin Nieto$^a$, Michael H. Holzscheiter$^b$,
and Slava G. Turyshev$^c$}\\

\vspace{2ex}
\end{center}

\begin{center}
$^a${\it Theoretical Division (MS-B285),
Los Alamos National Laboratory, \\
University of California, Los Alamos, NM 87545} \\[3pt]
$^b${\it Pbar Labs, LLC, 1601 Dove Street, Suite 170, Newport Beach, 
CA 92660} \\[3pt]
$^{c}${\it Jet Propulsion Laboratory, California Institute of Technology, \\
4800 Oak Grove Drive, Pasadena, CA 91109}
\end{center}
\vspace{1ex}

\begin{abstract}
To world-wide notice, in 2002 the ATHENA collaboration at CERN (in
Geneva, Switzerland) announced the creation of order 100,000 low
energy antihydrogen atoms.  Thus, the concept of using condensed
antihydrogen as a low-weight, powerful fuel (i.e., it produces a
thousand times more energy per unit weight of fuel than
fission/fusion) for very deep space missions (the Oort cloud and
beyond) had reached the realm of conceivability.  We briefly discuss
the history of antimatter research and focus on the technologies that
must be developed to allow 
a future use of controlled, condensed antihydrogen
for propulsion purposes. We emphasize that a dedicated antiproton
source (the main barrier to copious antihydrogen production) must be
built in the US, perhaps as a joint NASA/DOE/NIH project.  This is
because the only practical sources in the world are at CERN and the
proposed facility at GSI in Germany. We outline the scope and
magnitude of such a dedicated national facility and identify critical
project milestones. We estimate that, 
starting with the present level of knowledge
and multi-agency support, the goal of using antihydrogen for
propulsion purposes may be accomplished in $\sim50$ years.
\end{abstract}


\vspace{2ex}
\section{Introduction}

In this century, the development of missions  to deeper and deeper
space will become an ever increasing priority.
To complete a mission within a reasonable time frame,
even to the nearest extra-solar system objects of interest,
the Oort Cloud or the Alpha Centauri star system (4.3 light years
away), the velocity of the spacecraft needs to be high, up to more
than 10\% of the speed of light.  To achieve 
this one needs the highest energy-density fuel conceivable.  This
would be antimatter; a large amount of it and in a compact form. 

Antimatter can produce three orders of magnitude more energy per gram
than fission or fusion and ten orders of magnitude more energy than
the chemical reactions currently used for 
propulsion. As a result, it is a prime candidate for use in future
exploration beyond the solar system. It also is a candidate for future
missions to the edge of the solar system, which now
require on the order of 15-20 years after launch just to reach Pluto.

In this talk (MMN) we start with a 
quick review of both the discovery of and also our
understanding of antimatter (Section 2).  In Section 3 a description
is given of how cold antihydrogen was created in 2002.  We point
out why this is the only form of antimatter which is practical for
deep space propulsion.  We then take a side trip into current studies
using antiprotons for cancer therapy (Section 4).  
This side trip is important
as medical research may help with the funding necessary to yield large
amounts of antimatter.  In Section 5 we outline the trail we need to
break to obtain the dense antihydrogen that would be needed
for deep space travel.  We go on in Section 6 with a
discussion of what we can do now to start on this path, providing a
roadmap towards the goal.   Our conclusions follow.


\section{History of Antimatter}

It turns out that,
given quantum mechanics and special relativity, antimatter's existence
is a consequence \cite{r1}-\cite{3m}.
Although there are hints of the possible existence of antimatter in the
{\it strong reflection} solutions of special-relativity, the complete
break though came after Dirac discovered his relativistic equation for the
hydrogen atom \cite{r16}, whose solutions precisely agreed with the
observed energy levels.  

That is, this equation had four solutions, which could be interpreted
as those for particles with energy and internal spin properties
 \be 
\Psi_{Dirac} \sim \left\{
+E~\mathrm{spin~up},~+E~\mathrm{spin~down},~
-E~\mathrm{spin~down}, ~ - E~\mathrm{spin~up}\right\}. \label{e24}
\ee 
But the last two solutions had {\it negative energies}.  This led to a
huge controversy which was only resolved when Anderson discovered
the positron in 1932 \cite{r23,r23b}. This is an (anti)particle with
the same mass as but opposite electric charge as the electron.

Over the years, the antiproton, the antineutron, and, indeed with the
development of modern particle accelerators, all possible forms of
antimatter that can be detected have been detected.  We have come to
understand antimatter theoretically in terms of the $CPT$-Theorem of
modern field theory.        

In an intuitive form, the theorem says that if one were to take a
motion picture of a physical process and if one then were to
change the "charges" or ``internal quantum numbers" of the
particles in the movie ($C$),  run the
film backwards ($T$), and look at it in a mirror after rotating
oneself by 180$^\circ$
then one would not be able to tell the difference in the laws
of physics being seen.
Put another way, this theorem states that every particle has an
antiparticle with
{}
\begin{itemize}
\item[i)] the opposite electric charge,

\item[ii)] the opposite internal quantum numbers,

\item[iii)] the opposite magnetic moment,

\item[iv)] the same total lifetime, and

\item[v)] the same (inertial) mass. 
\end{itemize}
Although active searches continue for violations of this theorem, none
has been found.

Most importantly for us, if a particle and an
antiparticle collide they annihilate each other.  For example, if a
positron hits an electron, they turn into two high energy gamma rays,
each of energy of the rest mass of one particle, 511 keV. Stored
antimatter would be, by definition, the most powerful battery per unit
mass ever created.

Positrons (antielectrons) are now easily created in the laboratory
from $^{22}$Na sources and controlled in
Penning traps  \cite{penningtrap}.  With much more
difficulty (an efficiency of 1 part in 10$^{10}$) antiprotons are
created in high-energy accelerators.  At CERN in Geneva, Switzerland,
these antiprotons have been (again inefficiently) cooled and stored in
Penning traps for fundamental physics experiments.

However, these particles are by themselves not viable for antimatter
propulsion.  The storage volume must be small.  Charged antimatter 
is limited by the Brillouin density \cite{brillouin}
\be
n_0 = \frac{B^2}{2  \mu_0 m c^2}.
\ee
For antiprotons stored in a magnetic field using
today's technology, say 6 or even 25 T, 
this density would be around $10^{11}$ or $2 \times 10^{12}$
cm$^{-3}$, respectively.   (This number is itself orders of magnitude
higher than the highest antiproton density so far achieved, $\sim
10^6$ cm$^{-3}$ \cite{forever,ps200}.)  
Thus, charged antimatter is ruled out.
This leaves stable, neutral antimatter, i.e., antihydrogen.

Since, as we come to in the next section, cold antihydrogen has now 
been produced in the laboratory, it has been
argued Ref. \cite{nhp} that a fundamental science program needs to
be undertaken to manufacture and control dense antihydrogen, first in
the form of a cold dense gas or even a Bose-Einstein Condensate.  The
long range goal is to eventually
obtain condensed antihydrogen, either as a molecular superfluid, a
cluster ion, or as a diamagnetic solid.  This would allow a compact
source of antimatter to be used for deep-space propulsion.  But as many
have argued, its use would be tremendously powerful \cite{genta}.


\begin{figure}[h]
    \begin{center}
    \mbox{\epsfig{file=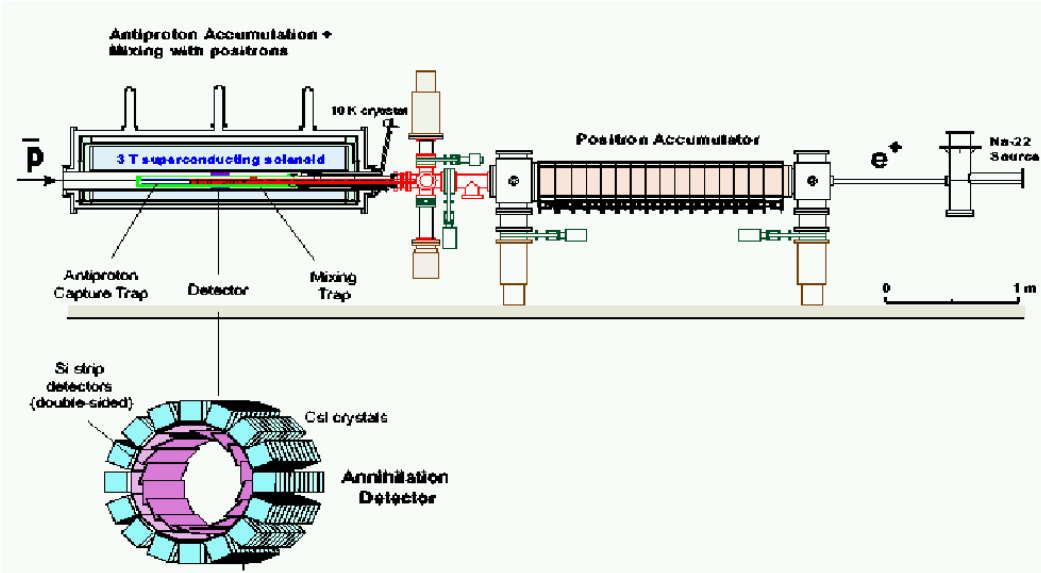,width=5.5in}}
    \end{center}
   \caption[General lay-out of the ATHENA experiment.]
   {\small
General lay-out of the ATHENA experiment \cite{NIM 2003}. Shown are the
positron accumulator and the main magnet system holding the
antiproton catching trap, the final positron storage trap, and the
recombination region. The antihydrogen detector surrounds the trap.
}
    \label{athena-lay-out}
 \end{figure}



\section{How Cold Antihydrogen was Created (2002)}

As positrons and antiprotons have been produced, then clearly then 
antihydrogen should also be able to be made.  But it
was not so easy. Until recently only  a few atoms of antihydrogen
had been produced at CERN \cite{HbarCERN} and at Fermilab \cite{HbarFNAL} in
high energy collisions. But these were
produced at relativistic speeds, much too fast to capture and
study.  But in late 2002, the ATHENA collaboration announced it
had produced the first low-energy antihydrogen atoms
(50,000 of them, later more) \cite{athenahbar}, using antiprotons which had 
been in a Penning trap that had  been extracted from
the AD (Antiproton Decelerator) at CERN. 
The excitement this produced was magnified by coverage in
the international press \cite{press}. 


\begin{figure}[h!]
    \begin{center}
    \mbox{\epsfig{file=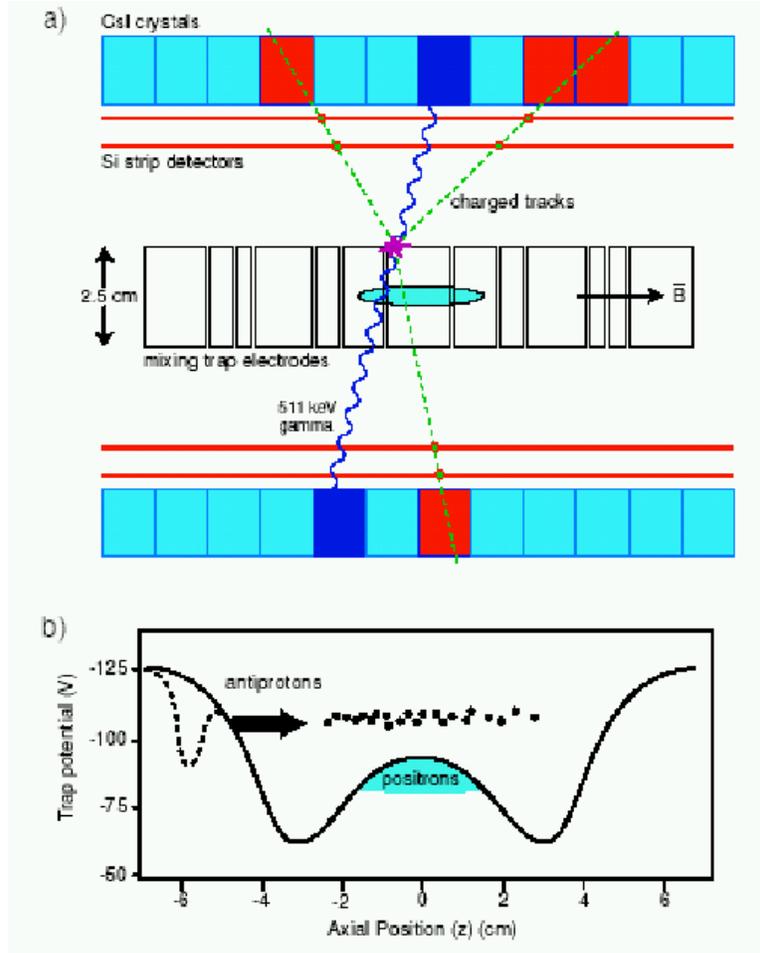,width=4in}}
    \end{center}
   \caption[Schematic of the central ATHENA apparatus and
trapping potential.]
   {\small
   \label{mhhc1}
Schematic of the central portion of the ATHENA apparatus and
the trapping potential used \cite{athenahbar}. 
(a) Section of the mixing trap and
detector showing the cylindrical electrodes and the position of
the positron cloud. A typical antihydrogen
annihilation event with the emission of three charged pions and a
pair of back-to-back 511 keV gamma-rays is shown. (b) The trapping
potential on axis is plotted along the length of the trap. The
dashed line shows the potential before the antiprotons and
positrons are mixed. 
}
\end{figure}


The general lay-out of the ATHENA experiment is shown in Figure
\ref{athena-lay-out}.  
The central portion of the ATHENA apparatus is
shown schematically in Figure \ref{mhhc1}(a), whilst the relevant trap
potentials are illustrated in Figure \ref{mhhc1}(b).

In each antiproton beam extraction, about $10^4$ antiprotons are
mixed with about $10^8$ positrons.   Once the low-energy antihydrogen
atoms are produced, they are neutral and hence no longer is bound in
the Penning trap configuration.
This means they are free to wonder in the direction of they momentum
after creation and they annihilate with normal
matter once they reach any matter in the trap, preferentially at the
walls.  The signal of an event
is the simultaneous (within 1 $\mu$s) detection of (i) two back to back
511 keV gamma rays (from the positron annihilating with an electron)
and (ii) about three charged pions (from the antiproton annihilating with a
nucleon) with the pions' momenta directions all
converging backwards to a single vertex point (to within a few mm)
which is on the line of the emitted photons.

\begin{figure}[h]
    \centering
    \noindent\mbox{\epsfig{file=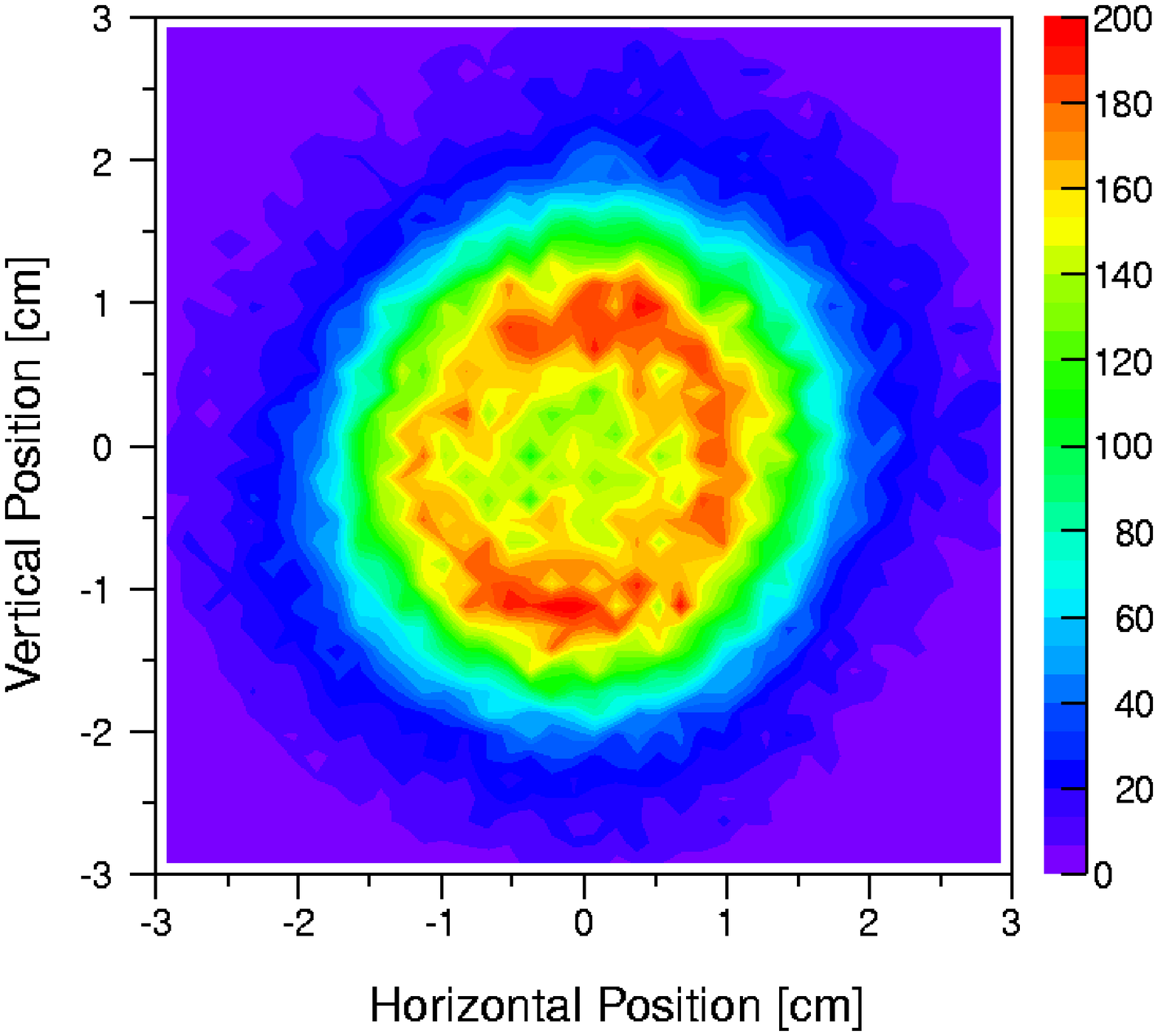,width=5.5 in}}
   \caption{Contour plot of the distribution (obtained by projecting
    onto the plane perpendicular to the magnetic field) of the vertex
    positions of reconstructed antihydrogen annihilation events from
    the ATHENA cold antihydrogen production experiment \cite{athenahbar}.
    \label{antihdie}}
\end{figure}

In Figure  \ref{antihdie} we shown the verification of the creation of
antihydrogen by the detection of the
annihilation products, preferentially on the walls of the trap.


Shortly after the ATHENA discovery the
ATRAP experiment also announced antihydrogen production 
\cite{atraphbarA,atraphbar}.
It is now the goal of these collaborations working at the AD to cool
these antihydrogen atoms even further, to confine them in possibly a
magnetic trap, and to perform experiments with them.

This completed a major
step on the road to the technology we are envisioning for
antimatter deep-space propulsion.


\section{Not-really-a-side-bar:  Antiproton Cancer Therapy}
\label{sec4}

Simultaneously with the antihydrogen experiments at CERN, the
low-energy antiproton beam from the AD was being used by the AD-4
collaboration to study the effect of antiprotons on living tissue 
as a precursor to possible cancer therapy.
An advantage of antiprotons over protons or
heavy ions is expected from the extra burst of annihilation energy
deposited at the stopping point (Bragg Peak).  By proper choice of the
beam energy this point can be located precisely inside the tumor
volume, would give a higher proportion of destruction to the cancer 
cells vs. the normal tissue the beam went through.

Preliminary results indicate this is true \cite{P324X} and
also that perhaps as few as $10^{10}$ antiprotons could treat a tumor
of size about 1 cm$^3$ \cite{AD-4unpub}.  Since present accelerators,
such as the former AC/AA combination at CERN or the future 
facility at GSI, produce on the order of $10^{14}$ antiprotons/year,
this therapy application has entered the realm of being a realistic 
possibility.  State-of-the-art modifications to
current accelerator designs could possibly produce a factor 100 more
antiprotons.  

Such  production would make antiproton therapy realizable, 
and the funds to produce such a source might be reasonably requested
from the NIH.  As emphasized in Section \ref{sec6.1},this could lead
to a symbiotic funding partnership with NASA. 
Indeed, discussions to pursue such a source for therapy 
purposes are already under
way.\footnote{A straw-man design is currently being studied by the
AD-4 collaboration \cite{P324X}.}


\section{The Route to Dense Antihydrogen and Deep Space Travel}

Long term storage of substantial amounts of antimatter must be
developed to enable space missions relying on antimatter-based
propulsion systems.   Although it is  clear that 
ultimately neutral antimatter must be used,  up to now, no valid
long-term storage concept for large quantities of antihydrogen has
been developed. On the other hand,
now that cold antihydrogen has been created, the next steps are to
to capture it and to cool it even further.  Designs for the first
goal are now being developed at CERN.  They concentrate on being able
to build a trap that trap the plasmas before combination and yet
also trap the neutral antihydrogen afterwards.  

This would be done
by surrounding the Penning trap configuration with a magnetic
quadrupole configuration, yielding a magnetic field 
\be
\mathbf{B} = B_0\left[\mathbf{\hat{z}} + \frac{(x\mathbf{\hat{x}} 
         - y\mathbf{\hat{y}})}{R_0}\right].
\ee
with a minimum at its center.
One would then use the magnetic dipole force on the antihydrogen atom 
\be
{\mathbf F}_{mag}= {\mathbf{\mu \cdot B}}
\ee
to trap the atoms in the so-called ``low-field seeking states.'' 
(The upper two states in the Hyperfine diagram for the ground state of
atomic hydrogen.)  

If this difficult work succeeds (and there appear to be no
matter-of-principle problems with it) the next goal will be to cool the
captured antihydrogen atoms to very low temperatures, perhaps using the
Lyman-alpha lasers that are being developed.

The first step in producing dense antihydrogen would be to produce
what has been done for hydrogen atoms, a Bose-Einstein Condensate
(BEC).\footnote{A BEC is a gaseous coherent quantum system, just as are
superfluid helium or superconducting currents.}  BEC 
confinement of neutral spin-polarized hydrogen atoms at densities up to
$5 \times 10^{15}$ cm$^{-3}$ has been demonstrated. \cite{BEC}, 
which is orders of magnitude more dense than the Brillouin
storage density limit for charged 
antiprotons. To make a BEC of
antihydrogen would be an individual step, where one could learn the
techniques of controlling a relatively large amount of antihydrogen.
One also would have to overcome the problem of the antihydrogen 
transitioning out of the confined states \cite{nhp}.
The clear ultimate goal would be to make 
very dense antihydrogen, in the form of clusters or solids (perhaps
stored diamagnetically).

At present the wasteful method of resonant evaporative cooling is used
to achieve the temperatures and densities needed to form a hydrogen
BEC. But the development of lasers for 
direct and efficient cooling of hydrogen atoms has now just
started. Efficient laser cooling of hydrogen will revolutionize the
methodology of forming, controlling, and studying
hydrogen Bose-Einstein Condensates.  These studies can all be done
with ordinary matter, in preparation for having more copious amounts
of antihydrogen available. 

If the envisioned progress comes to fruition, laser cooling could then
be used in an attempt to efficiently make an antihydrogen BEC.
An antihydrogen BEC would be an important  step down a path that could
eventually lead to even more dense antihydrogen molecules, liquids,
solids. and cluster ions 
\cite{silvera}-\cite{young}.  Indeed, since one might expect the next
stage to be going from controlled ultra-cold (below 50 $\mu$K) BEC
hydrogen atoms to controlled hydrogen
molecules, it is heartening that there is evidence of a
hydrogen-molecule superfluid with a critical temperature of 0.15 K
\cite{grebenev}.  Since the triple point of hydrogen is at 
13.8 K, a potential path to denser condensed antimatter becomes more
interesting. 


\section{What Can We Do Now?}

A space-certified storage system for neutral antimatter
will not be obtained from a linear extrapolation of heretofore existing
technologies. Rather, it  requires a series
scientific and/or technological breakthroughs. While breakthroughs can
never be predicted, they typically will not happen 
without the definition of a strong need and the challenge presented to
the scientific community by a truly ambitious goal. 

Meanwhile many of
the underlying issues can be addressed
with both the modest supply of antimatter available at this time at
accelerator centers world wide and with the limited means to store the
particles. The technological and 
scientific knowledge gained in these tests will enable us to lay out a
path into the future of antimatter-based propulsion systems. 

However, the most important item is the need for a dedicated
low-energy antiproton source in the United States.


\subsection{A dedicated Low-Energy Antiproton Source in the USA}
\label{sec6.1}

The biggest obstacle to producing copious antihydrogen is the dearth
of low-energy antiproton production.  As stated, it presently is a very
inefficient process and is done only in Europe.
At present the only source of low-energy antiprotons is at CERN. 
It is hoped that the AD facility at CERN will keep
running until perhaps the end of this decade. Then a newly proposed
facility, FLAIR (A facility for Low-energy Antiproton and Ion
Research), will hopefully be built at the GSI accelerator center in Germany. 
it could yield $10^{12}$ {\it Low Energy} antiprotons per year.

But the US needs a facility so it can realisticly pursue the ultimate
goal of copious production of antiprotons leading to copious
antihydrogen.  It is only with a viable facility that studies can be
done that will lead to the necessary break-through technology
needed for more efficient antiproton production.

The communities to do this, perhaps as as consortium, are there.  The
DOE physics community would like such a facility to continue fundamental
symmetry studies on antimatter and also to test gravity \cite{3m}.
As pointed out in Section \ref{sec4}, the work on antiproton  
therapy would lead to NIH interest in this facility.
NASA would have an interest for deep-space flight.
There are also other communities that would have an interest; space
reactor teams, RTG builders, radiation physicians and physicists, and
nuclear and particle experimentalists.  A NASA/DOE/NIH consortium to
build a dedicated facility would be a natural.


\subsection{A Roadmap To Antihydrogen Propulsion}
\label{sec6.2}

Knowing the cost of acquiring the technological capabilities
needed to produce large quantities of antihydrogen atoms,
to store them for long periods, and to use them for propulsion purposes in
space is, of course, very important.  Given our present technological
level, our estimates are that: 

$\bullet$ It would take about 5 years and $\sim$ 0.5 B\$ to build a source.

$\bullet$ It would take about the same time and money  more to develop
antihydrogen handling technologies.  

During all this time effort would be given to developing the new
antiproton production technology that is needed.
  Current antiproton production rates are low. While clever techniques
can enhance these rates by several order of magnitude and quantities
sufficient for advanced concepts can be 
produced given enough economic and political pressure onto the few
available sources, a real breakthrough can only come through continued
interest and research in this area. A good
analogy is the comparison between a light bulb and a laser. In both
cases light is produced, but in one system through thermal heating of
a material and in the other through 
coherent processes. Antiprotons are currently produced by heating a
metal target with a primary proton beam. This is a direct analogy to
the light bulb --- we are still awaiting the
invention of a `laser-equivalent' for the production of particles of
antimatter.

$\bullet$ A GUESS is that 10-20 years more would be needed for this.

This would be the make or break point.  If after 30 years one did not have a
new antiproton production technology, then the effort would be
abandoned as far as deep-space propulsion is concerned, although not
for the other applications.  But with success, 

$\bullet$ A BIGGER GUESS is that it would take 10-20 more years to
develop a real system.

Note that much of the technology will  be standard in the sense that
the power transfer from antimatter to thrust has long been a 
problem of interest \cite{rand,forward},
as well as that of obtaining thrust from other nuclear 
mechanisms \cite{MICF,ACMF}.

So, we are talking of about 50 B\$ over 50 years.  That period is like the time
from vacuum-tube computers to the microchip processors of today.  It
would be a viable time frame - if it works.  But most importantly,
antimatter science has now advanced to the point where antimatter
technology has left the realm of science fiction and has reached the
first stages of reality.\footnote{Indeed, if one considered positron
emission tomography (PET), the reality arrived some time ago.}


\section{Conclusions}

The road we have described is challenging both scientifically and
technologically.   Enormous scientific and technological barriers have
to be overcome.  But the potential intellectual and societal rewards,
even along the way, are enormous.

Antimatter-matter annihilation is one of the prime candidates to
achieve the high specific impulse  i) desired for the challenging
missions of exploring the Heliopause and  visiting the Oort Cloud, and
ii) needed if we plan to attempt a 
rendezvous with the nearest star systems. While no clear pathway to
the necessary technologies exists, experimental development in the normal
matter world of laboratory-sized research
equipment can help us to  reach these most ambitious goals, {\it IF} 
we simultaneously embark on constructing
a dedicated low-energy antiproton facility

It behooves us to now embark on extensive, serious work on the
possibilities that are before us. To achieve them quickly it is
necessary to set ourselves in motion now.  


\section*{Acknowledgements}
MMN and MHH acknowledge the support of the United
States Department of Energy, partially under contract W-7405-ENG-36
(MMN). The work of SGT was performed at the
Jet Propulsion Laboratory, California Institute of Technology, under
contract with the  National Aeronautics and Space Administration.



\end{document}